\documentclass[11pt,a4paper,fleqn]{article}

\usepackage{amsmath,amssymb,amsthm,enumerate}

{\theoremstyle{definition}

\newtheorem*{note*}{Note}
}

\begin{document}
%\hspace{12 cm}
%IPM/P-2004/004
\par\noindent {\LARGE\bf
Biot-Savart helicity versus physical helicity:\\ A topological description of ideal flows\\

\par}

{\vspace{5mm}\par\noindent {\it
TALIYA SAHIHI~$^\dag$ HOMAYOON ESHRAGHI~$^\star$ \par\vspace{2mm}\par}

{\vspace{2mm}\par\noindent \it
Physics Department,
Iran University of Science and Technology (IUST), Tehran, Iran}

{\vspace{2mm}\par\noindent {$\phantom{\dag}$~\rm $^\dag$\,\, }{\it
taliya\_sahihi@iust.ac.ir} \par}

{\vspace{2mm}\par\noindent {$\phantom{\dag}$~\rm $^\star$\,\, }{\it
eshraghi@iust.ac.ir
} \par}

{\vspace{8mm}\par\noindent\hspace*{10mm}\parbox{140mm}{\small

For an isentropic (thus compressible) flow, fluid trajectories are considered as orbits of a family of one parameter, smooth, orientation-preserving and nonsingular diffeomorphisms on a compact and smooth-boundary domain in the Euclidian 3-space which necessarily preserve a finite measure, later interpreted as the fluid mass. Under such diffeomorphisms the Biot-Savart helicity of the pushforward of a divergence-free and tangent to the boundary vector field is proved to be conserved and since these circumstances present an isentropic flow, the conservation of the ``Biot-Savart helicity" is established for such flows. On the other hand, the well known helicity conservation in ideal flows which here we call it ``physical helicity" is found to be an independent constant with respect to the Biot-Savart helicity. The difference between these two helicities reflects some topological features of the domain as well as the velocity and vorticity fields which is discussed and is shown for simply connected domains the two helicities coincide.     The energy variation of the vorticity field is shown to be formally the same as for the incompressible flow obtained before. For fluid domains consisting of several disjoint solid tori, at each time, the harmonic knot subspace of smooth vector fields on the fluid domain is found to have two independent base sets with a special type of orthogonality between these two bases by which a topological description of the vortex and velocity fields depending on the helicity difference is achieved since this difference is shown to depend only on the harmonic knot parts of velocity, vorticity and its Biot-Savart vector field. For an ideal MHD flow three independent constant helicities are reviewed while the helicity of magnetic potential is generalized for non-simply connected domains by inserting a special harmonic knot field in the dynamics of the magnetic potential. It is proved that the harmonic knot part of the vorticity in hydrodynamics and the magnetic field in MHD is presented by constant coefficients (fluxes) when expanded in terms of one of the time dependent base functions.

}\par\vspace{6mm}}

\vspace{5mm} \hspace{2mm} Key words: Topological flows, Helicity, Harmonic knots.

\hspace{2mm} Mathematics Subject Classification (2010): 76B47, 37B99.

%%%%%%%%%%%%%%%%%%%%%%%%%%%%%%%%%%%%%%%%%%%%%%%%%%%%%%%%%%%%%%%%%%%%%%%%%%%%%%%%%%%%%%%%%%%%%%
\section{Introduction}

Fluid dynamics as a famous sample for dynamical systems  reveals a few topological aspects. Even before the discovery and full construction of topology, some structural properties being conserved in fluid flows for long times were observed. Such structural properties finally turned out to be satisfactorily explained via topological objects \cite{moffatt1}. The nonlinearity as well as the specific type of hydrodynamical governing equations implied a number of open problems mainly within the context of geometric analysis although there are some aspects explained via pure topology, for example, what types of topological features are preserved by the fluid advection? Or in other words, what are the topological invariants in the fluid motion?

Although a complete answer has not yet been addressed, but very useful invariants have been discovered among which helicity in vortex and magnetic
field dynamics is of particular importance. Historically, topological considerations for the vortex dynamics were initiated by Helmholtz in 1858 \cite{helm} and later by Kelvin in 1869 \cite{kelvin}. They found vortex lines are frozen into the fluid motion governed by the Euler equation and moreover Kelvin detected the persistence of knot and links of vortex lines. Two decades later in 1893 Poincar\'{e} reconsidered and generalized these results within a geometric dynamical system approach \cite{poincare}. However, a more elaborate and rigorous understanding of the subject was achieved after 1958 when Woltjer and Chandrasekhar through the study of magnetic lines in Crab Nebula recognized the conservation of magnetic helicity. In addition, Woltjer proved that the minimization of the magnetic energy under the helicity conservation is satisfied by the Beltrami field \cite{chandra}-\cite{woltjer2}. In spite of fundamental differences between magnetic and vortex fields, the frozenness of both fields into the fluid motion, causes the possibility to define similar invariant helicities. Nevertheless, vortex helicity was not discovered immediately after Woltjer`s work, but instead, Moffatt in 1969 discovered the similarity between magnetic and vortex invariants and introduced the name ``helicity" for the first time \cite{moffat2}. To complete the historical survey, let us mention the role of Moreau in this story that according to Moffatt`s declaration \cite{moffat3}, Moreau in 1961 understood the helicity of two linked vortex tubes. He published these results in a brief communication of Comptes Rendus de l'acad\'{e}mie des Sciences.

Moffatt and probably Woltjer were aware of the closed relation between this helicity and Gauss's integral formula for the linking number of two
separate closed curves \cite{gauss} which may lead to the writhing number defined for a single closed curve indicating the measure of its complexity \cite{caluga1}-\cite{caluga3}. Indeed, helicity is a generalization of linking and writhing numbers to a $C^1$ vector field, especially when magnetic or vortex lines shrink into a finite number of closed tubes, the helicity reduces to a linear combination of all linking and writhing numbers. A more rigorous mathematical generalization of writhing and linking numbers to  the concept of helicity for $C^\infty$ vector fields was done by Arnold in 1973 (1974) \cite{arnold1}. It is not known whether Arnold knew anything about Moffatt's results because he used the name ``asymptotic Hopf invariant" for the helicity based on a theorem proved by him that this asymptotic Hopf invariant reduces to an integral obtained by the Biot-Savart formula \cite{arnold1}-\cite{gambaudo1}. The helicity introduced by Moffatt is the integral of the inner product of the physical fluid velocity and vorticity which we call it the ``physical helicity". In contrast, a special helicity is usually used in mathematics literature defined only on some domains in $\mathbb{R}^3$ which is obtained via the Biot-Savart formula and let us call it the ``Biot-Savart helicity". At first Arnold introduced a helicity for a certain kind of divergence-free vector fields on a closed orientable 3-manifold and showed this helicity is preserved under a volume-preserving and orientation-preserving diffeomorphism. In the case when the manifold is a 3D smooth compact submanifold of $\mathbb{R}^3$ with smooth boundary, this helicity preservation splits into several cases which are not considered here \cite{cantarella1}. Then Arnold showed this helicity is the same as the asymptotic Hopf invariant for simply connected manifolds \cite{arnold1}-\cite{khesin}. Again for a connected smooth domain in $\mathbb{R}^3$, the asymptotic Hopf invariant reduces to the Biot-Savart helicity \cite{gambaudo1}.

Another important result of Arnold is that the field energy (that is the integral of the squared of the divergence-free vector field) has lower bound proportional to its helicity. Arnold generalized the above helicity preservation to the helicity conservation under the action of a family of volume-preserving and orientation-preserving diffeomorphisms obtained from a real parameter which are differentiable with respect to this parameter. It then follows that the field energy takes a local minimum when the vector field possesses definite conditions, for example considering the magnetic field as this divergence-free vector field, Arnold's result means that the magnetic field in a perfectly conducting plasma has the minimum energy when the plasma is at equilibrium \cite{arnold1,arnold2}. As mentioned before, for a compact 3-D submanifold of $\mathbb{R}^3$ with smooth boundary, the asymptotic Hopf invariant coincides with the Biot-Savart helicity and the lower bound for the field energy becomes more understandable and easier to work with. This task was done by Cantarella \emph{et al} in 2000-2001 \cite{cantarella1}-\cite{cantarella3}. In those works, the Biot-Savart helicity conservation under the action of a volume-preserving and orientation-preserving family of smooth diffeomorphisms implies an incompressible flow field carrying the domain and the divergence-free vector field defined on it.

Observation on ideal fluid dynamics show another possibility for the frozenness of vector lines in fluid flows leading to the conservation of the physical helicity (as defined earlier) and therefore a natural question arises whether a Biot-Savart helicity is conserved along a compressible
flow. This issue is addressed in the present note resulting in a positive answer for an isentropic (consequently compressible) fluid. Subsequently the energy variation is derived to be formally equal to the incompressible case \cite{cantarella1} and new conditions for the energy minimization is studied. The next fundamental enquiry investigated here is the topological difference between the two conserved helicities, namely the physical and the Biot-Savart helicities for an isentropic flow in any compact domain in $\mathbb{R}^3$ with smooth boundary. Specially when the fluid domain consists of several disjoint solid multiple tori, a simple algebraic structure is found for harmonic knot fields which helps a better understanding of the difference between two helicities. Relatively similar situation exists in the dynamics of magnetic fields in ideal conducting fluids leading to three distinct types of conserved helicities; the Biot-Savart, the magnetic potential and the cross helicities among which the magnetic potential helicity is generalized here to be more adequate for non-simply connected fluid domains. The above main problems accompanied by a few minor consequences construct the contents of the present article according to the following structure. The next section (Sec.~2) is devoted to prove the conservation of the Biot-Savart helicity carried by a compressible smooth flow assigned with a family of orientation-preserving and nonsingular smooth diffeomorphisms and necessarily preserving a finite measure which models the mass of the real fluid. This conservation probably relates to the helicity preservation under a diffeomorphism which generally preserves a measure although this may be quite different from the result presented here \cite{contreras}-\cite{cantarella4}. The energy variation for the vector field is derived in Sec.~3 followed by the discussion on the conditions for the energy minimization. In Sec.~4 a brief survey on isentropic fluid equations and especially the frozenness of vortex lines are reviewed to justify the Biot-Savart helicity conservation derived in Sec.~2. Section 4 is continued with paying attention to the difference between the two helicities which exists only for non-simply connected fluid domains. A finer treatment requires employing the Hodge decomposition theorem on compact domains in $\mathbb{R}^3$ with smooth boundary \cite{cantarella1}-\cite{cantarella4} mentioned briefly here. It is shown that the harmonic knot subspace of vector fields on domains consisting of disjoint solid tori is spanned alternatively by two different bases of functions with a special kind of orthogonality which is found to be important for the interpretation of the helicity difference. In Sec.~5 ideal MHD equations are cited to drive and describe the three different types of helicity as mentioned above. Finally Sec.~6 summarizes all results with some conclusive remarks.

%%%%%%%%%%%%%%%%%%%%%%%%%%%%%%%%%%%%%%%%%%%%%%%%%%%%%%%%%%%%%%%%%%%%%%%%%%%%%%%%%%%%%%%%%%%%
\section{Writhing number and helicity}

The writhing number of a smooth closed curve (knot) in the 3-space is a numerical invariant up to isotopy which represents the net number of times that the curve winds around itself. This quantity can be calculated by using the formula which
was introduced by C\u{a}lug\u{a}reanu
\begin{equation}
Wr(C)=\frac{1}{4\pi} \int_{C\times C}(\frac{dx}{ds}\times \frac{dy}{dt})\cdot\frac{x-y}{|x-y|^3}ds dt.
\end{equation}
where $C$ is the image of the knot and $s$ and $t$ are arc length parameters along the curve at the same scale. Since the curves are considered to model a wide variety of objects in several phenomena, the writhing number can be used as a topological criterion and a dynamical invariant in studying those phenomena. For example, in molecular biology the writhing number describes how many times the axis of a DNA helix has crossed itself.

Now suppose a smooth vector field $V$ on $\Omega$ as a compact domain in $\mathbb{R}^3$ with smooth boundary $\partial\Omega$. Each closed integral curve of the vector field is like a space curve and therefore the writhing number can be generalized to a standard measure for the behavior of the vector field lines if all field lines are closed, the case usually happens when the vector field is divergence-free. In this sense, Woltjer introduced a new quantity later called helicity by Moffatt, which is defined through the formula
\begin{equation}
\mathcal{H}(V)=\frac{1}{4\pi} \int_{\Omega\times\Omega}V(x)\times V(y)\cdot\frac{x-y}{|x-y|^3}d^3xd^3y.
\end{equation}
Similar to the well known Biot-Savart law in magneto-statics, one can construct from $V$ a new smooth vector field $BS(V)$ on $\mathbb{R}^3$ as
\begin{equation}
BS(V)(y)=\frac{1}{4\pi} \int_{\Omega}V(x)\times \frac{y-x}{|y-x|^3}d^3x,
\end{equation}
Restricting the domain of $BS(V)$ implies the linear operator (Biot-Savart operator)
\begin{equation}
BS:\Gamma(\Omega) \rightarrow \Gamma(\Omega),
\end{equation}
where $\Gamma(\Omega)$ denotes for the space of smooth vector fields on $\Omega$. Comparing Eqs.~(2) and (3) leads to
\begin{equation}
\mathcal{H}(V)=\int_{\Omega}V(x)\cdot BS(V)(x) d^3x.
\end{equation}
Let $\Gamma(\Omega)$ be considered as an $\textit{L}^2$-space endowed with the inner product as
\begin{equation}
\langle V\mid W\rangle=\int_{\Omega}V(x)\cdot W(x)d^3x,
\end{equation}
by which the helicity is expressed as
\begin{equation}
\mathcal{H}(V)=\langle V\mid BS(V)\rangle.
\end{equation}

In further considerations in this note, vector fields are assumed to be divergence-free and tangent to the boundary $\partial\Omega$ unless indicated otherwise. In this case the Biot-Savart operator is nothing but an inverse of the curl operator.
\begin{equation}
\nabla\times BS(V)=V.
\end{equation}
Obviously $BS(V)$ is not necessarily divergence-free and tangent to $\partial\Omega$. However using the Hodge decomposition theorem, it is possible to modify $BS(V)$ to satisfy these conditions \cite{cantarella2}. This modification is not used in this paper.

The writhing number which was defined for a single knot can be easily generalized to the ``linking number" of two separate curves $C_1$ and $C_2$:
\begin{equation}\nonumber
Link(C_{1},C_{2})=\frac{1}{4\pi} \int_{C_{1}\times C_{2}}(\frac{dx}{ds}\times \frac{dy}{dt})\cdot\frac{x-y}{|x-y|^3}ds dt.
\end{equation}
Indeed for a single curve, the writhing number is the linking number of the curve with itself (self linking number). Further generalization deals with two smooth divergence-free vector fields $V_1$ and $V_2$ on $\Omega_1$ and $\Omega_2$ respectively, each tangent to the boundary of its relevant domain. Equation (2) is then generalized to the definition of a mutual (cross) helicity
\begin{equation}\nonumber
\mathcal{H}(V_1,V_2)=\frac{1}{4\pi} \int_{\Omega_1\times\Omega_2}V_1(x)\times V_2(y)\cdot\frac{x-y}{|x-y|^3}d^3xd^3y,
\end{equation}
which is obviously symmetric in $V_1$ and $V_2$. A direct substitution of Eq.~(3) gives
\begin{equation}\nonumber
\mathcal{H}(V_1,V_2)=\langle V_1\mid BS(V_2)\rangle=\langle V_2\mid BS(V_1)\rangle=\mathcal{H}(V_2,V_1).
\end{equation}

Now suppose $\Omega_{0}$ is a domain in 3-space ($\mathbb{R}^3$) with smooth boundary and let
$\{h_{t}:\Omega_{0}\rightarrow\Omega_{t}\mid t\in\mathbb{R} , \Omega_{t}\subset \mathbb{R}^3\}$ be a family of one parameter, orientation-preserving, smooth and nonsingular diffeomorphisms such that $h_{0}$ is the identity. For each $x\in\Omega_{0}$, let $y_{t}=h_{t}(x)$ and so $d^{3}y_{t}=J_{t}d^{3}x$ where $J_{t}$ is the Jacobian of $h_{t}$. Since $h_{t}$ was assumed to be smooth and nonsingular, $J_{t}$ neither changes sign and nor becomes infinite and since $J_{0}=1$ ($h_{0}$ is the identity) it is clear that $J_{t}$ remains positive and finite for all $t\in\mathbb{R}$. This property admit the definition of a finite measure $\mu$, preserved under $h_{t}$ for each $t$ as follows. If the volume element $d^{3}x$ in $\Omega_{0}$ contains $d\mu$ we can define a density $\lambda_{0}(x)$ for this measure such that $d\mu=\lambda_{0}(x)d^{3}x$. Since $d\mu$ must be preserved, we have $d\mu=\lambda_{t}d^{3}y_{t}=J_{t}\lambda_{t}d^{3}x$ and so
\begin{equation}
\lambda_{t}(x)= \frac{\lambda_{0}(x)}{J_{t}(x)}.
\end{equation}
The smooth diffeomorphism $h_{t}$ induces a smooth vector field $W_{t}$ on $\Omega_{t}$ defined as
\begin{equation}
W_{t}(y_t)= \frac{\partial}{\partial t}h_{t}(x)|_{x=(h_{t})^{-1}(y_{t})}.
\end{equation}
It is not difficult to prove that \cite{Chorin}
\begin{equation}
\frac{D J_{t}}{D t}:=\frac{\partial}{\partial t}J_{t}(x)=J_{t}\nabla_{y_{t}}\cdot W_{t},
\end{equation}
where $\nabla_{y_{t}}\cdot$ denote the divergence with respect to $y_{t}$.
The two different notations $\frac{D}{D t}$ and $\frac{\partial}{\partial t}$ for the same concept emphasize the two expression of $J_{t}$: If $J_{t}$ is considered as a function of $x$ and $t$, we use $\frac{\partial}{\partial t}$ and when $J_{t}$ is expressed in terms of $y_{t}$ and $t$ (since $x=(h_{t})^{-1}(y_{t})$) the notation $\frac{D}{D t}$ is used. This means that
\begin{equation}
\frac{D}{D t}:=(\frac{\partial}{\partial t})_{y_{t}}+W_{t}\cdot\nabla_{y_{t}},
\end{equation}
where $(\frac{\partial}{\partial t})_{y_{t}}$ means the time derivative at constant $y_{t}$
Equations (9)-(12) yield
\begin{equation}
(\frac{\partial \lambda_{t}}{\partial t})_{y_{t}}+\nabla_{y_{t}}\cdot(\lambda_{t}W_{t})=0,
\end{equation}
which is exactly the continuity equation for the preserved measure.

In the next step let $\omega_{0}$ be a smooth divergence-free vector field on $\Omega_{0}$ and tangent to the boundary of $\Omega_{0}$. This admits the definition of $\omega_{t}$ on $\Omega_{t}$ as
\begin{equation}\
\omega_{t}=\lambda_{t}(h_{t})_{*}(\frac{\omega_{0}}{\lambda_{0}}),
\end{equation}
where $(h_{t})_{*}$ denotes the pushforward of $\omega_{0}$. In order to construct a helicity for $\omega_{t}$ on $\Omega_{t}$ one must first assure that $\omega_{t}$ is divergence-free and tangent to $\partial\Omega_{t}$. This is achieved through the following lemma.\vspace{2.5mm}\\
\textbf{Lemma 1.} Under the hypothesis and definitions of Eq.~(14), $\omega_{t}(y_{t},t)$ is divergence-free and tangent to the boundary of $\Omega_{t}$, that is
\begin{equation}
\forall \;t, y_{t}\;\;\;\;\;\;\;\nabla_{y_{t}}\cdot\omega_{t}=0,\;\;\;\;\;\;\text{and}\;\;\;\;\;\;\omega_{t}\cdot n=0, \;\text{on}\;\partial\Omega_{t}.
\end{equation}\vspace{2.5mm}\\
\emph{Proof:} Equation (14) means that
\begin{equation}
\frac{\omega_{t}}{\lambda_{t}}=(\frac{\omega_{0}}{\lambda_{0}}\cdot\nabla_{x})y_{t},
\end{equation}
while its time derivative $\frac{D}{Dt}$, since $y_{t}=h_{t}(x)$, Eq.~(10) implies
\begin{equation}
\frac{D}{Dt}(\frac{\omega_{t}}{\lambda_{t}})=(\frac{\omega_{0}(x)}{\lambda_{0}(x)}\cdot\nabla_{x})W_{t}.
\end{equation}
Due to the hypothesis, is a diffeomorphism, and from Eq.~(16) one finds
\begin{equation}
\frac{\omega_{t}}{\lambda_{t}}\cdot\nabla_{y_{t}}=\sum_{i=1}^{3}\frac{\omega_{ti}}{\lambda_{t}}\frac{\partial}{\partial y_{ti}}=
\sum_{i,j=1}^{3}\frac{\omega_{0j}}{\lambda_{0}}\frac{\partial y_{ti}}{\partial x_{j}}\frac{\partial}{\partial y_{ti}}=
\sum_{j=1}^{3}\frac{\omega_{0j}}{\lambda_{0}}\frac{\partial}{\partial x_{j}}=\frac{\omega_{0}}{\lambda_{0}}\cdot\nabla_{x}.
\end{equation}
substituting (18) into (17) leads to
\begin{equation}
\frac{D}{Dt}(\frac{\omega_{t}}{\lambda_{t}})=(\frac{\omega_{t}}{\lambda_{t}}\cdot\nabla_{y_{t}})W_{t}.
\end{equation}
Employing Eq.~(12) to express $\frac{D}{Dt}$ and using Eq.~(13) and the vector identity $\nabla\times(A\times B)=(B\cdot\nabla)A-(A\cdot\nabla)B+(\nabla\cdot B)A-(\nabla\cdot A)B$ finally Eq.~(19) converts to
\begin{equation}
\frac{\partial\omega_{t}}{\partial t}=\nabla_{x}\times(W_{t}\times\omega_{t})-W_{t}(\nabla_{y_{t}}\cdot\omega_{t}).
\end{equation}
Now one can take the divergence with respect to $y_{t}$ from the above equation and recall Eq.~(12) to obtain
\begin{equation}\nonumber
\frac{D}{Dt}(\nabla_{y_{t}}\cdot\omega_{t})+(\nabla_{y_{t}}\cdot\omega_{t})(\nabla_{y_{t}}\cdot W_{t})=0,
\end{equation}
and again from Eq.~(13) we find
\begin{equation}\nonumber
\lambda_{t}\frac{D}{Dt}(\nabla_{y_{t}}\cdot\omega_{t})-(\nabla_{y_{t}}\cdot\omega_{t})\frac{D\lambda_{t}}{Dt}=0,
\end{equation}
and since $\lambda_{t}$ never vanishes or diverges, we can divide this equation by $\lambda_{t}^{2}$ and obtain
\begin{equation}\nonumber
\frac{\nabla_{y_{t}}\cdot\omega_{t}}{\lambda_{t}}=\frac{\nabla_{x}\cdot\omega_{0}}{\lambda_{0}}=\text{constant}.
\end{equation}
Finally we remember that $\omega_{0}$ was assumed to be divergence-free and this proves that $\omega_{t}$  is divergence-free.

To prove that $\omega_{t}$ is tangent to $\partial\Omega_{t}$ we use the relation $d\mu=\lambda_{t}d^{3}y_{t}=\lambda_{0}d^{3}x$ and since the volume element $d^{3}y_{t}$ is being advected by the velocity $W_{t}$ we may write $d^{3}y_{t}=ds_{t}\cdot dl_{t}$ where $ds_{t}$ and $dl_{t}$ are respectively the surface and line elements constructing $d^{3}y_{t}$ both being pushforwarded by $h_{t}$. The relation $\frac{D}{Dt}(d\mu)=0$ where $d\mu=\lambda_{t}d^{3}y_{t}=\lambda_{t}ds_{t}\cdot dl_{t}$ implies that
\begin{equation}\
\frac{D}{Dt}(\lambda_{t}ds_{ti})=-\lambda_{t}\frac{\partial W_{tj}}{\partial y_{ti}}ds_{tj},\;\;\;\;\;\;\;(i,j=1,2,3)
\end{equation}
in which the relation
\begin{equation}\
\frac{D}{Dt}(dl_{t})=(dl_{t}\cdot\nabla_{y_{t}})W_{t},
\end{equation}
has been used. From Eqs. (19) and (21) it follows that
\begin{equation}\
\frac{D}{Dt}(\omega_{t}\cdot ds_{t})=0,
\end{equation}
consequently $\omega_{t}\cdot ds_{t}=\omega_{0}\cdot ds_{0}$ and since $\omega_{0}$ was assumed to be tangent to $\partial\Omega_{0}$, at any time $t$, $\omega_{t}$ is also tangent to $\partial\Omega_{t}$.\hspace{0.25cm}$\blacksquare$\vspace{.25cm}\\
From this lemma $\omega_{t}$ is divergence-free and Eq.~(20) reduces to
\begin{equation}\
(\frac{\partial\omega_{t}}{\partial t})\mid_{y_{t}}=\nabla_{y_t}\times(W_{t}\times\omega_{t}).
\end{equation}
The preceding lemma allows one to define the helicity $\mathcal{H}(\omega_{t})$ according to Eq.~(2) or Eq.~(5) over $\Omega_{t}$. This helicity is constant as follows.\vspace{2.5mm}\\
\textbf{Theorem 1.} The helicity $\mathcal{H}(\omega_{t})$  defined by Eq.~(5) is constant with respect to $t$ that is
\begin{equation}\
\frac{d}{dt}\mathcal{H}(\omega_{t})=0.
\end{equation}
\vspace{.25cm}\vspace{2.5mm}\\
\emph{Proof:} Starting from Eq.~(5) we change the end coordinates from $y_{t}$ to $x$ through $(h_{t})^{-1}$ and write $d^{3}y_{t}=J_{t}d^{3}x$. Then substitute $J_{t}$ from Eq.~(9) to obtain
\begin{equation}
\mathcal{H}(\omega_{t})=\int_{\Omega_{0}}\frac{\omega_{t}}{\lambda_{t}}\cdot BS(\omega_{t})\lambda_{0}(x) d^3x,
\end{equation}
where $\omega_{t}$ and $\lambda_{t}$ must be expressed in terms of $x$ and $t$. Thus
\begin{equation}\nonumber
\frac{d}{dt}\mathcal{H}(\omega_{t})=\int_{\Omega_{0}}\frac{D}{D_{t}}(\frac{\omega_{t}}{\lambda_{t}}\cdot BS(\omega_{t}))\lambda_{0} d^3x.
\end{equation}
Applying Eq.~(19) and using the relation (12) after changing coordinates to $y_{t}$ again yields
\begin{equation}
\frac{d}{dt}\mathcal{H}(\omega_{t})=\int_{\Omega_{t}}[(\omega_{t}\cdot\nabla_{y_{t}})W_{t}]\cdot BS(\omega_{t})d^3y_{t}+\int_{\Omega_{t}}\omega_{t}\cdot\frac{\partial BS(\omega_{t})}{\partial t}d^3y_{t}+\int_{\Omega_{t}}\omega_{t}\cdot[(W_{t}\cdot\nabla_{y_{t}})BS(\omega_{t})]d^3y_{t}.
\end{equation}
In the next step we show that the first and third terms on the right side of Eq.~(27) cancel each other. To do this, let us notice to the first term and use the fact that $\omega_{t}$ is divergence-free:
\begin{equation}\nonumber
\int_{\Omega_{t}}[(\omega_{t}\cdot\nabla_{y_{t}})W_{t}]\cdot BS(\omega_{t})d^3y_{t}=
\int_{\Omega_{t}}(BS(\omega_{t}))_{i}\nabla_{y_{t}}\cdot(W_{ti}\cdot\omega_{t})d^3y_{t}=
\end{equation}
\begin{equation}\nonumber
\int_{\Omega_{t}}\nabla_{y_{t}}\cdot[(BS(\omega_{t}))_{i}(W_{ti}\cdot\omega_{t})]d^3y_{t}
-\int_{\Omega_{t}}W_{ti}(\omega_{t}\cdot\nabla_{y_{t}})(BS(\omega_{t}))_{i}d^3y_{t},
\end{equation}
where the summation over repeated index $i$ is implicit. The first term converts to a surface integral which since $\omega_{t}$ is tangent to $\partial\Omega_{t}$ (by Lemma 1) vanishes. Thus the first and third terms of Eq.~(27) become
\begin{equation}\nonumber
\int_{\Omega_{t}}\omega_{tj}W_{ti}(\frac{\partial (BS(\omega_{t}))_{j}}{\partial y_{ti}}-\frac{\partial (BS(\omega_{t}))_{i}}{\partial y_{tj}}) d^3y_{t}=\int_{\Omega_{t}}W_{ti}\omega_{tj}(\nabla_{y_{t}}\times BS(\omega_{t}))_{k}\epsilon_{ijk}d^3y_{t}=
\end{equation}
\begin{equation}\nonumber
\int_{\Omega_{t}}W_{ti}\omega_{tj}\omega_{tk}\epsilon_{ijk}d^3y_{t}=0,
\end{equation}
where $\epsilon_{ijk}$ is the Levi-Civita symbol. Thus, we find
\begin{equation}
\frac{d}{dt}\mathcal{H}(\omega_{t})=\int_{\Omega_{t}}\omega_{t}\cdot (\frac{\partial BS(\omega_{t})}{\partial t})_{y_{t}}d^3y_{t},
\end{equation}
According to Eq.~(3) we write
\begin{equation}\nonumber
(\frac{\partial BS(\omega_{t})}{\partial t})_{y_{t}}=\frac{1}{4\pi}\frac{\partial}{\partial t}\int_{\Omega_{t}}\omega_{t}(y'_{t},t)\times\frac{ y_{t}-y'_{t}}{\mid y_{t}-y'_{t}\mid^{3}}d^3y'_{t}=
\end{equation}
\begin{equation}\nonumber
\frac{1}{4\pi}\int_{\Omega_{t}}\frac{\partial\omega_{t}(y'_{t},t)}{\partial t}\times \frac{y_{t}-y'_{t}}{\mid y_{t}-y'_{t}\mid^{3}}d^3y'_{t}
+\frac{1}{4\pi}\int_{\partial\Omega_{t}}\omega_{t}(y'_{t},t)\times\frac{y_{t}-y'_{t}}{\mid y_{t}-y'_{t}\mid^{3}}(W_{t}\cdot ds_{t}(y'_t)),
\end{equation}
We then substitute the above relation into Eq.~(28) and change the order of integration to find
\begin{equation}
\frac{d}{dt}\mathcal{H}(\omega_{t})=\int_{\Omega_{t}}\frac{\partial\omega_{t}}{\partial t}\cdot BS(\omega_{t})d^3y'_{t}+
\int_{\partial\Omega_{t}}\omega_{t}\cdot BS(\omega_{t})(W_{t}\cdot ds_{t}(y'_{t})),
\end{equation}
in which Eq.~(3) again has been applied.From Eq.~(24) the first term of Eq.~(29) reduces to
\begin{equation}\nonumber
\int_{\Omega_{t}}\nabla_{y'_{t}}\times(W_{t}\times\omega_{t})\cdot BS(\omega_{t})d^3y'_{t}
+\int_{\Omega_{t}}\nabla_{y'_{t}}\cdot[(W_{t}\times\omega_{t})\times BS(\omega_{t})]d^3y'_{t}
-\int_{\Omega_{t}}(W_{t}\times\omega_{t})\cdot\nabla_{y'_{t}}\times BS(\omega_{t})d^3y'_{t},
\end{equation}
The first term in the above expression becomes a surface integral and the second term vanishes according to Eq.~(8). Thus, the first term on the right hand side of Eq.~(29) converts to
\begin{equation}\nonumber
-\int_{\partial\Omega_{t}}\omega_{t}\cdot BS(\omega_{t})(W_{t}\cdot ds_{t}(y'_t))+
\int_{\partial\Omega_{t}} W_{t}\cdot BS(\omega_{t}) (\omega_{t}\cdot ds_{t}(y'_t)).
\end{equation}
The first term in the above cancels the second term in the right hand side of Eq.~(29) while the second term in the above vanishes since $\omega_{t}$ is tangent to $\partial\Omega_{t}$ and this completes the the proof.\hspace{0.25cm}$\blacksquare$

%%%%%%%%%%%%%%%%%%%%%%%%%%%%%%%%%%%%%%%%%%%%%%%%%%%%%%%%%%%%%%%%%%%%%%%%%%%%%%%%%%%%%%%%%%%%
\section{Energy variation of a vector field}
By the energy of a vector field is meant the $\textit{L}^{2}$ inner product of vector field with itself.
Thus the energy of $\omega_t$ on the domain $\Omega_t$ is defined as
\begin{equation}
E(\omega_t)= \langle\omega_t\mid \omega_t\rangle=\int_{\Omega_t}\omega_t^2 d^3y_t.
\end{equation}

Although the name energy to this quantity mainly comes from historical considerations but in some special physical cases this quantity is really a physical energy. For example if $\omega_t$ denotes for a magnetic field frozen into the plasma flow or when it denotes the velocity of an incompressible fluid. However if $\omega_t$ stands for the vorticity of a fluid flow $E(\omega_t)$ is called the ``enstrophy''. Therefore this quantity is relatively important at least in physical applications and it pays to have a quick look and its variation through the following simple theorem.\vspace{.25cm}\\
\textbf{Theorem 2.} The rate of the energy change of $\omega_t$ with respect to $t$ is given by:
\begin{equation}
\frac{d}{dt}E(\omega_{t})= 2\langle W_{t}\mid\omega_{t}\times(\nabla\times\omega_{t})\rangle-\int_{\partial\Omega_{t}}\omega_{t}^{2}(W_{t}\cdot ds_{t}).
\end{equation}\vspace{.25cm}\\
\emph{Proof:} The proof is almost similar to the previous argument provided the use of $\lambda_td^3y_t=\lambda_0d^3x$.
\begin{equation}
\frac{d}{dt}E(\omega_{t})=\int_{\Omega_{0}}\left[\frac{D\omega_{t}}{Dt}\cdot\frac{\omega_{t}}{\lambda_{t}}
+\omega_{t}\cdot\frac{D}{Dt}(\frac{\omega_{t}}{\lambda_{t}})\right]\lambda_{0}d^{3}x.
\end{equation}
Utilizing Eqs.~(12), (19) and (24) gives rise to
\begin{equation}
\frac{d}{dt}E(\omega_{t})=\int_{\omega_{t}}\omega_{t}\cdot\left[\nabla_{y_{t}}\times(W_{t}\times\omega_{t})
+(W_{t}\cdot\nabla_{y_{t}})\omega_{t}+(\omega_{t}\cdot\nabla_{y_{t}})W_{t}\right]d^{3}y_{t}.
\end{equation}
Then we use the vector identity $\nabla(A\cdot B)=(A\cdot\nabla)B+(B\cdot\nabla)A+A\times(\nabla\times B)+B\times(\nabla\times A)$ to rewrite the second and third terms and obtain
\begin{equation}
\frac{d}{dt}E(\omega_{t})=\int_{\Omega_{t}}\omega_{t}\cdot[\nabla_{y_{t}}\times(W_{t}\times\omega_{t})
+\nabla(W_{t}\cdot\omega_{t})-W_{t}\times(\nabla_{y_{t}}\times\omega_{t})-\omega_{t}\times(\nabla\times W_{t})]d^{3}y_{t}.
\end{equation}
The last term directly vanishes while the second term because $\omega_{t}$ is divergence--free converts to a surface integral which also vanishes since $\omega_{t}$ is tangent to the boundary. Inserting the identity $\nabla\cdot(A\times B)=B\cdot(\nabla\times A)-A\cdot(\nabla\times B)$ with $A=W_{t}\times\omega_{t}$ and $B=\omega_{t}$ in the first term yields
\begin{equation}
\frac{d}{dt}E(\omega_{t})=\int_{\Omega_{t}}[\nabla_{y_{t}}\cdot((W_{t}\times\omega_{t})\times\omega_{t})
+2(\nabla_{y_{t}}\times\omega_{t})\cdot(W_{t}\times\omega_{t})]d^{3}y_{t}.
\end{equation}
The first integral converts to the surface integral which by expanding the integrand through the BAC-CAB rule and again using the fact that $\omega_t$ is tangent to $\partial\Omega_t$ finally Eq.~(31) follows.\hspace{0.25cm}$\blacksquare$\vspace{.25cm}

Notice that, this energy variation formally coincides with that obtained in Reference \cite{cantarella1} for the case of volume-preserving diffeomorphisms.
An interesting special case happens when $\omega_t$ is the eigenvector of the curl operator that is
\begin{equation}
\nabla_{y_{t}}\times\omega_{t}=\xi\omega_{t},
\end{equation}
where $\xi$ is a constant. In this case the energy change rate reduces to
\begin{equation}
\frac{d}{dt}E(\omega_{t})=-\int_{\partial\Omega_{t}}\omega_{t}^{2}(W_{t}\cdot ds_{t}).
\end{equation}
On the other hand Eq.~(36) leads to the following integral condition
\begin{equation}
0=\int_{\Omega_{t}}\omega_t\times\nabla_{y_{t}}\times\omega_{t}d^{3}y_{t}=\frac{1}{2}\int_{\partial\Omega_{t}}\omega_{t}^{2}ds_{t},
\end{equation}
where we have used the identity $A\times\nabla\times A=\nabla (A^2/2)-(A\cdot\nabla)A$ and Lemma 1. According to Eq.~(37) in this special case the energy may have an extremum at $t$ which may happen if $W_{t}\cdot ds_{t}$ changes sign on $\partial\Omega_{t}$. Therefore in cases where condition (36) holds and the domain $\partial\Omega_{t}$ is everywhere expanding or everywhere collapsing the energy can not have any extremum.

Unfortunately it is not possible to prove that condition (36) is preserved for all $t$ and even it is a very difficult task to look for a sufficient condition for its preservation. So all one can assume is that condition (36) may occur only at some special values of $t$. Indeed condition (36) with the aid of Eq.~(8) yields
\begin{equation}
BS(\omega_t)=\xi^{-1}\omega_t+\nabla_{y_{t}}\phi+\zeta,
\end{equation}
where $\phi$ is a scalar field and $\zeta$ is a divergence-free and curl-free (harmonic knot) vector field tangent to $\partial\Omega_t$ both defined on $\Omega_t$. This decomposition is in fact the Hodge decomposition (discussed in the next section) and when $\Omega_t$ is simply connected $\zeta\equiv 0$. Therefore according to definitions (5) and (30), condition (39) gives
\begin{equation}
\mathcal{H}(\omega_t)=\xi^{-1}E(\omega_t)+\langle\omega_t\mid\zeta\rangle=\xi^{-1}E(\omega_t)+\int_{\partial\Omega_{t}}(BS(\omega_t)\times\zeta)\cdot ds_t,
\end{equation}
where for the last equality we have used the identity $\nabla\cdot(A\times B)=B\cdot(\nabla\times A)-A\cdot(\nabla\times B)$ together with Eq.~(5) and the fact that $\zeta$ is curl-free. If $\zeta$ vanishes or more generally, as will be seen in the next section, if $\zeta$ is orthogonal to the harmonic part of $\omega_t$ the helicity at that time becomes proportional to the energy, the case studied by Arnold and others which is important in finding upper bounds for helicity \cite{arnold1,cantarella1,cantarella3}.

%%%%%%%%%%%%%%%%%%%%%%%%%%%%%%%%%%%%%%%%%%%%%%%%%%%%%%%%%%%%%%%%%%%%%%%%%%%%%%%%%%%%%%%%%%%%
\section{Helicity in hydrodynamics}

As mentioned in Sec.~1 for a hydrodynamical system in physics Moffatt and Moreau had already discovered a helicity which is constant by the fluid motion. We referred to this helicity as ``physical helicity" and want to show that physical helicity is in general different from the helicity introduced in the second section named the ``Biot-Savart helicity". Consider an ideal, isentropic, compressible fluid filling a domain $\Omega_0$ in $\mathbb{R}^3$ with smooth boundary. Let $u_0(x)$ and $\omega_{0}=\nabla_{x}\times u_{0}$ are the initial fluid velocity and vorticity vector fields respectively such that $\omega_0$ is tangent to the boundary of $\Omega_0$. Assume the fluid particle trajectories are presented by the family of smooth orientation-preserving diffeomorphisms $\{ h_{t}:\Omega_{0}\rightarrow\Omega_{t}\mid t\in\mathbb{R} , \Omega_{t}\subset \mathbb{R}^3\}$ such that at any time $t$,$y_t=h_t(x)\in\Omega_t$ is the place of the particle trajectory at this time initiating from the place $x$ at $t=0$. Following the notations and concepts of Sec.~2 we have
\begin{equation}
u_{t}= \frac{\partial}{\partial t}h_{t}(x)|_{x=(h_{t})^{-1}(y_{t})}.
\end{equation}
As discussed before, this family of smooth diffeomorphisms preserves a measure which for the fluid it can be considered to be the fluid mass $m$. The fluid mass density $\rho_t$ plays the role of $\lambda_t$. From this the continuity equation follows similar to Eq.~(13)
\begin{equation}
(\frac{\partial \rho_{t}}{\partial t})_{y_{t}}+\nabla_{y_{t}}\cdot(\rho_{t}u_{t})=0.
\end{equation}
In an isentropic ideal flow the fluid velocity $u_t$ must satisfy the following equation of motion
\begin{equation}
\frac{Du_t}{Dt}=-\nabla_{y_{t}} w_t,
\end{equation}
where $w_t$ is the thermodynamical enthalpy per unit mass of the fluid and $\frac{D}{D t}$ is defined in the same sense of Eq.~(12) in which $W_t$ is replaced by $u_t$. In the above equation we apply the vector identity written just after Eq.~(38) and then take the curl and used the identity written just after Eq.~(19) to obtain
\begin{equation}
(\frac{\partial\omega_{t}}{\partial t})_{y_{t}}=\nabla_{y_t}\times(u_{t}\times\omega_{t}),
\end{equation}
where $\omega_t=\nabla_{y_t}\times u_t$ is the fluid vorticity. This equation is exactly similar to Eq.~(24), therefore the proof of Lemma 1 immediately leads to the pushforwardness (frozenness) of the vorticity by the fluid advection:
\begin{equation}
\frac{D}{Dt}(\frac{\omega_{t}}{\rho_{t}})=(\frac{\omega_{t}}{\rho_{t}}\cdot\nabla_{y_{t}})u_{t},
\end{equation}
whose solution is
\begin{equation}
\frac{\omega_{t}}{\rho_{t}}=(\frac{\omega_{0}}{\rho_{0}}\cdot\nabla_{x})y_{t}.
\end{equation}
We emphasize that only dynamical equations of the fluid leads to this fact that $\frac{\omega_t}{\rho_t}$ is the pushforward of $\frac{\omega_0}{\rho_0}$ along the fluid particle trajectories. This enables us to define the Biot-Savart helicity $\mathcal{H}_{BS}(\omega_t)$  exactly the same as given in Theorem 1:
\begin{equation}
\mathcal{H}_{BS}(\omega_t)=\int_{\Omega_t}\omega_t\cdot BS(\omega_t)d^{3}y_{t},
\end{equation}
which is a constant of the motion directly by Theorem 1.

On the other hand the physical helicity
\begin{equation}
\mathcal{H}_{Ph}(u_t)=\int_{\Omega_t}u_t\cdot \nabla_{y_{t}}\times u_t d^{3}y_{t}=\int_{\Omega_t}u_t\cdot\omega_{t} d^{3}y_{t},
\end{equation}
is another constant of the motion whose proof is given below. Just like in Eq.~(26) we write
\begin{equation}\nonumber
\frac{d}{dt}\mathcal{H}_{Ph}(u_t)=\int_{\Omega_t}\left(\frac{D}{Dt}\frac{\omega_t}{\rho_t}\right)\cdot u_t \rho_t d^{3}y_t+\int_{\Omega_t}\frac{\omega_t}{\rho_t}\cdot\frac{Du_t}{Dt}\rho_t d^{3}y_t,
\end{equation}
in which the relation $dm=\rho_td^{3}y_t$ has been used. Using Eqs.~(43) and (45) and converting the above volume integrals to surface integrals together with the fact $\omega_t$ is tangent to the boundary finally proves the claim.

Hence, at each time $t$ there exist two different fundamental helicities on $\Omega_t$ both constant at all times and thus one can define $\Delta \mathcal{H}$ to be the difference of these two helicities:
\begin{equation}\nonumber
\Delta \mathcal{H}(u_t)=\mathcal{H}_{Ph}(u_t)-\mathcal{H}_{BS}(\nabla_{y_{t}}\times u_t)=\int_{\Omega_t}[u_t-BS(\nabla_{y_{t}}\times u_t)]\cdot \nabla_{y_{t}}\times u_t d^{3}y_{t}
\end{equation}
\begin{equation}
=\int_{\Omega_t}[u_t-BS(\omega_t)]\cdot\omega_{t} d^{3}y_{t}.
\end{equation}
A natural question arises: what does this difference mean in general? To answer this question mathematically it is better to review the properties of the Biot-Savart operator and use the description of smooth vector fields through the Hodge decomposition theorem cited below. Before starting this procedure  we draw the reader's attention to this important fact that the Biot-Savart helicity is a constant whose nature comes only from the frozenness property of the vorticity in the fluid motion and it is a  direct consequence of the curl of the Euler equation that is Eq.~(44) or (45). Thus some important physical information is killed when taking this curl, for example if in the righthand side of the Euler equation one can add any other curl-free vector field without changing Eq.~(44) and therefore without changing the Biot-Savart helicity. As will be seen in the next section the magnetic field is also frozen into the ideal plasma flow which gives a Biot-Savart helicity as well.  On the other hand the constancy of the physical helicity as seen above was originated directly from the mechanics (Newton's second law) and thermodynamics of fluid elements (Eq.~(43)) and though this helicity is enough reach of physical information. However the difference between these two helicities in some aspects maybe even more enlightening because the less important part of this constant (Biot-Savart helicity) is being removed. As will be seen in Theorem 4 there is another important physical constant called the Kelvin circulation theorem which is also a direct consequence of the Euler equation (43) and not its curl.
As mentioned in the Introduction $BS(\omega_{t})$ can be extended to the whole of $\mathbb{R}^3$ and also it can be easily seen that
\begin{equation}
(BS(\omega_{t}))(y_{t})=\nabla_{y_{t}}\times (P(\omega_{t}))(y_{t}),
\end{equation}
where
\begin{equation}
(P(\omega_{t}))(y_{t})=\frac{1}{4\pi}\int_{\Omega_{t}}\frac{\omega_{t}(y'_{t})}{\mid y_{t}-y'_{t}\mid}d^{3}y'_{t}\,\,\,\,\,\,\,\text{and}\,\,\,\,\,\,\,\nabla_{y_{t}}\cdot P(\omega_{t})=0,
\end{equation}
which immediately implies that $BS(\omega_t)$ is divergence-free. It is not difficult to see that $\nabla_{y_{t}}\times (BS(\omega_{t}))(y_{t})$ equals $\omega_{t}(y_{t})$ if $y_{t}\in\Omega_t$ and $0$ otherwise. Consider a closed curve (a knot) $C$ in $\Omega_t$ bounding a disk $D\subset\mathbb{R}^{3}$. Stokes' theorem implies that
\begin{equation}
\oint_{C}BS(\omega_{t})\cdot dl=\int_{D\cap \Omega_t}\omega_{t}\cdot dS.
\end{equation}
This means that if the interior of the disk $D$ lies totally outside $\Omega_{t}$ (necessarily $C\subset\partial\Omega_{t}$), the above integral vanishes and thus $BS(\Omega_t)$ is an Amperian vector field. We will not go further in discussing the properties of the Biot-Savart operator and refer the interested reader to References \cite{cantarella1,cantarella2}.

Assume the velocity field $u_t$ which was originally defined on $\Omega_t$ can be extended to the whole of $\mathbb{R}^3$ but not necessarily smooth or even continuous outside $\Omega_t$. Such extension is possible and definitely not unique and one may consider it with a compact support including $\Omega_t$. However it is better to impose the restriction that the $u_t$ and its first derivatives be integrable and quadratic integrable on whole volumes, surfaces and curves in $\mathbb{R}^3$. This situation is very common in fluid dynamics specially when the domain $\Omega_t$ is embedded in the total region occupied by the fluid, that is, the fluid exists outside $\Omega_t$ in a larger domain while $\Omega_t$ always contains the same definite fluid particles. Under these circumstances the line integral of Eq.~(52) for the extended fluid velocity becomes

\begin{equation}
\oint_{C}u_t\cdot dl=\int_{D}\nabla_{y_{t}}\times u_{t}\cdot dS.
\end{equation}
This equation obviously indicates the role of extended fluid velocity which directly affects the line integral of $u_t$ on any closed curve (knot) inside $\Omega_t$ while such possibility was absent for $BS(\omega_t)$. Finally a more precise description follows from the use of the Hodge decomposition theorem. This well known theorem generally deals with algebraic structures in the space of smooth differential forms on compact oriented smooth Riemannian manifolds \cite{warner} and even more generally the complex structures and mixed structures in algebraic geometry. However specially for compact domains in the Euclidian 3-space with piecewise smooth boundary it takes a very nice form \cite{cantarella1,cantarella2,cantarella4}:
\vspace{4mm}\\
\textbf{Theorem (Hodge decomposition).} Let $\Omega$ be a compact domain in $\mathbb{R}^3$ with piecewise smooth boundary and $\Gamma(\Omega)$ be the infinite dimensional space of smooth vector fields on $\Omega$, then, this vector space uniquely splits into the direct sum of mutually orthogonal (in the $\emph{L}^{2}$ product according to Eq.~(6)) subspaces as follows:
\begin{equation}
\Gamma(\Omega)= FK(\Omega)\oplus HK(\Omega)\oplus CG(\Omega)\oplus HG(\Omega)\oplus GG(\Omega),
\end{equation}
where
\begin{equation}\nonumber
FK(\Omega)=\text{Fluxless knots}=\{V\in \Gamma(\Omega)\mid \nabla\cdot V=0, V\cdot n=0, \text{ all interior fluxes=0}\},
\end{equation}
\begin{equation}\nonumber
HK(\Omega)=\text{Harmonic knots}=\{V\in \Gamma(\Omega)\mid \nabla\cdot V=0, \nabla\times V=0, V\cdot n=0\},
\end{equation}
\begin{equation}\nonumber
CG(\Omega)=\text{Curly gradients}=\{V\in \Gamma(\Omega)\mid V=\nabla\varphi, \nabla\cdot V=0,\text{ all boundary fluxes=0}\},
\end{equation}
\begin{equation}\nonumber
HG(\Omega)=\text{Harmonic gradients}=\{V\in \Gamma(\Omega)\mid V=\nabla\varphi, \nabla\cdot V=0,\text{ locally constant on }\partial\Omega\},
\end{equation}
\begin{equation}\nonumber
GG(\Omega)=\text{Grounded gradients}=\{V\in \Gamma(\Omega)\mid V=\nabla\varphi,\varphi\mid_{\partial\Omega}=0\}.
\end{equation}
In addition
\begin{equation}
ker(curl)= HK(\Omega)\oplus CG(\Omega)\oplus HG(\Omega)\oplus GG(\Omega),
\end{equation}
\begin{equation}
image(curl)=FK\oplus HK\oplus CG,
\end{equation}
\begin{equation}\nonumber
image(grad)=CG\oplus HG\oplus GG,
\end{equation}
\begin{equation}\nonumber
ker(div)=FK\oplus HK\oplus CG\oplus HG.
\end{equation}\vspace{2.5mm}\\
Only two of these five subspaces namely $HK(\Omega)$ and $HG(\Omega)$ (together constructing the harmonic subspace) have definitely finite dimensions and the remaining subspaces are not generally finite dimensional. These finite dimensions are related to homology groups of the domain according to following isomorphisms.
\begin{equation}
HK(\Omega)\cong H_{1}(\Omega;\mathbb{R})\cong H_{2}(\Omega,\partial\Omega;\mathbb{R})\cong \mathbb{R}^{\text{total genus of all components of } \partial\Omega},
\end{equation}
\begin{equation}
HG(\Omega)\cong H_{2}(\Omega;\mathbb{R})\cong H_{1}(\Omega,\partial\Omega;\mathbb{R})\cong \mathbb{R}^{\sharp(\partial\Omega)-\sharp(\Omega)}.
\end{equation}
where $\sharp$ denotes for the number of components. The second isomorphism in each of the above two equations comes from the Poincar\'{e} duality. A detailed proof for the last isomorphisms for an arbitrary compact smooth domain $\Omega$ is presented in \cite{cantarella4}. An alternative proof is presented here but for simplicity we assume that $\Omega$ consists of several connected components each of which is a solid torus with an arbitrary genus. The advantage of this restriction is to find easily an orthogonality condition for subspace $HK(\Omega)$. Of course, each handlebody of these solid tori may be knotted in itself or different handles can have link in each other such complexities do not affect our discussion. A similar discussion may be considered for subspace $HG$ which is not cited here since it does not take part in dealing with helicity.

Corresponding to each component of $\Omega$ (or $\partial\Omega$) with genus $g$ there exist exactly $g$ nontrivial loops in $H_{1}(\Omega,\mathbb{R})$ each rounding exactly one handlebody. Let $C_{1},\ldots,C_{n}$ be such nontrivial loops in $\Omega$ each with the least complexity (as a knot) where $n$ is the sum of all genuses of components of $\partial\Omega$. We emphasize that since each handlebody of each torus component of $\partial\Omega$ may be knotted in itself and each $C_i$ is isotopic with the longitude circle of its containing handle, this loop definitely has the same type of knot. Similarly these loops link in each other exactly like the corresponding handles. According to the Seifert theorem for each knot $C_i$ there exists a connected oriented surface (Seifert surface) in $\mathbb{R}^{3}$ whose boundary is exactly $C_i$ and each $C_i$ is either a simple circle (unknot) or is a knot with the minimal complexity i.e. parallel to the corresponding handlebody. In the former case the Seifert surface is a disk necessarily not contained in $\Omega$ because otherwise the loop $C_i$ is trivial in $\Omega$. In the latter case since $C_i$ is parallel to the longitude knotted handle, its Seifert surface again must exceed that handle. Hence, in any case the Seifert surface of each $C_i$ is not a subset of $\Omega$. For each $C_i$ consider the cross section $\Sigma_i$ of its corresponding handle, transverse to it and properly embedded in $\Omega$ ($\partial\Sigma_i\subset\partial\Omega$) such that cutting $\Omega$ across these surfaces produces a simply connected domain.

For any $\omega\in HK(\Omega)$  it is possible to define
\begin{equation}
\kappa_i=\oint_{C_i}\omega\cdot dl,\,\,\,\,\,\,\,\,\,\,\Phi_i=\int_{\Sigma_i}\omega\cdot dS,
\end{equation}
in which the orientation of $\Sigma_i$ and $C_i$ are compatible. It must be emphasized that since the Seifert surface of each $C_i$ is not contained in $\Omega$ the first integral by Stokes' theorem may become different from zero in general. We show that $\Omega$ is uniquely determined if each circulation $\kappa_i$ is given, or is uniquely determined if each flux $\Phi_i$ is given. Assume $\omega_1,\omega_2\in HK(\Omega)$ both have exactly the same set of circulations ($\kappa_1,\ldots ,\kappa_n$). Then for the vector field $\omega=\omega_2-\omega_1$ all circulations are zero and for the simply connected domain obtained by cutting $\Omega$ across all $\Sigma_i$'s, we have a potential $\psi$ such that $\omega=\nabla\psi$. For example for each solid torus one may choose a point $x_0$ in the central part such that it is the intersection of all handle loops and after cutting $\Omega$ across $\Sigma_i$'s, this component becomes simply connected and one can define $\psi(x)=\int_{x_0}^{x}\omega\cdot dl$ in which the integration does not depend on the path. Each cross section $\Sigma_i$ can isotopically coincide with a surface of constant potential $\psi_{0i}$ which is perpendicular to $\partial\Omega$. Now instead of cutting $\Omega$ across any $\Sigma_i$ (which now is a surface of constant potential), remove a tiny neighborhood of it such that the potential $\psi$ takes the values $\psi_{0i}-\epsilon$ and $\psi_{0i}-\kappa_i+\epsilon$ on both sides $\Sigma_{i}^{-}$ and $\Sigma_{i}^{+}$ of this neighborhood respectively for sufficiently small $\epsilon$. Let us call the domain obtained by removing all these neighborhoods $\Omega_{\epsilon}$ and therefore $\Omega_0=\Omega-\bigcup_{i}\Sigma_i$. We thus have
\begin{equation}
\int_{\Omega}\omega^{2}d^{3}x=\lim_{\epsilon\rightarrow 0}\int_{\Omega_{\epsilon}}\mid\nabla\psi\mid^{2}d^{3}x.
\end{equation}
Since $\nabla^{2}\psi=0$, the last integral can be converted to the surface integral which since $\omega$ is tangent to $\partial\Omega$ it reduces to
\begin{equation}
\lim_{\epsilon\rightarrow 0}\sum_{i=1}^{n}\{\Phi_{i}(\psi_{0i}-\epsilon)-\Phi_{i}(\psi_{0i}-\kappa_{i}+\epsilon)\}=\sum_{i=1}^{n}\kappa_{i}\Phi_{i},
\end{equation}
and since for $\omega$ all circulations vanish we find $\omega=\nabla\psi=0$ which means that $\omega_1=\omega_2$. If on the other hand $\omega_1$ and $\omega_2$ have exactly the same set of fluxes ($\Phi_1,\ldots ,\Phi_n$) then for $\omega=\omega_2-\omega_1$ all $\Phi_i$'s are zero and again by Eqs.~(60) and (61) $\omega=0$ and $\omega_1=\omega_2$ and this completes the proof of the claimed uniqueness.

Therefore we can consider two different bases $f_1,\ldots,f_n$ and $l_1,\ldots,l_n$ for $HK(\Omega)$ such that for each $f_i$ we have $\Phi_j=\delta_{ij}$ and for each $l_i$ we have $\kappa_{j}=\delta_{ij}$. Moreover, a kind of orthogonality between these two bases is discussed here. For any $\epsilon$ let $\psi_{i}^{(l)}$ and $\psi_{j}^{(f)}$ denote for the potential for $l_i$ and $f_j$ on $\Omega_{\epsilon}$ respectively. Similar to Eqs.~(60) and (61) we can write
\begin{equation}
\langle l_i\mid f_j\rangle=\lim_{\epsilon\rightarrow 0}\int_{\Omega_{\epsilon}}\nabla\psi_{i}^{(l)}\cdot\nabla\psi_{j}^{(f)}d^{3}x,
\end{equation}
which after converting to the surface integral we find
\begin{equation}
\langle l_i\mid f_j\rangle=\sum_{k=1}^{n}(k\text{th circulation of }l_i)(k\text{th flux of }f_j)=0.
\end{equation}
Therefore an arbitrary $\omega\in HK(\Omega)$ has two different expansions:
\begin{equation}
\omega=\sum_{k=1}^{n}\langle\omega\mid l_k\rangle f_k=\sum_{k=1}^{n}\langle\omega\mid f_k\rangle l_k,
\end{equation}
thus
\begin{equation}
\langle\omega\mid f_k\rangle=\kappa_k,\,\,\,\,\,\,\,\,\,\,\langle\omega\mid l_k\rangle=\Phi_k.
\end{equation}
This equation for subspace $HK(\Omega)$  introduces an identity operator
\begin{equation}
\textrm{Id}_{HK(\Omega)}=\sum_{k=1}^{n}l_kf_k=\sum_{k=1}^{n}f_kl_k,
\end{equation}
which acts according to $\emph{L}^2$ product. This immediately yields
\begin{equation}
\langle\omega_1\mid \omega_2\rangle=\sum_{k=1}^{n}(k\text{th flux of }\omega_1)(k\text{th circulation of }\omega_2)=\sum_{k=1}^{n}(k\text{th flux of }\omega_2)(k\text{th circulation of }\omega_1).
\end{equation}
We saw each set $\{\kappa_1,\ldots,\kappa_n\}$ determines exactly one vector field $\omega\in HK(\Omega)$ and this vector field determines exactly one set $\{\Phi_1,\ldots,\Phi_n\}$ and this one to one correspondence is indeed an isomorphism between the two $n$-dimensional Euclidian spaces. The first space is clearly isomorphic to $H_{1}(\Omega;\mathbb{R})$ whose set of generators is $\{C_1,\ldots,C_n\}$. On the other hand $H_2(\Omega,\partial\Omega;\mathbb{R})$ is freely generated by $\{\Sigma_1,\ldots,\Sigma_n\}$ because the boundary of each $\Sigma_i$ is in $\partial\Omega$ and $\partial\Sigma_i\cap\partial\Omega$ does not bound a subsurface of $\partial\Omega$ which together with $\Sigma_i$ bound a 3-dimensional simplicial part. This argument must be supported by the isomorphism between singular and simplicial homology.
Summarizing the above discussion we proved the following theorem\vspace{.25cm}\\
\textbf{Theorem 3.}
If $\Omega$ is a disjoint union of solid multiple tori then the subspace $HK(\Omega)$ is given according to Eq.~(57) and moreover, there are two different bases for this subspace which satisfy orthogonality conditions as presented in Eq.~(63) with the relevant inner product structure through Eqs.~(64)-(67).\vspace{.25cm}

Let apply the above theorems for the fluid domain $\Omega_t$ which from now assume to be a disjoint union of solid tori (some of them may be balls in general). Therefore at each time $t$ we have the loops $C_{1t},\ldots,C_{nt}$ and cross sections $\Sigma_{1t},\ldots,\Sigma_{nt}$ and the corresponding bases $\{l_{1t},\ldots,l_{nt}\}$ and $\{f_{1t},\ldots,f_{nt}\}$ with the circulations $\kappa_{1t},\ldots,\kappa_{nt}$ and fluxes $\Phi_{1t},\ldots,\Phi_{nt}$ for each member of $HK(\Omega_t)$. There is another physical invariant not yet used called the Kelvin circulation by which we can obtain a nice representation of the $HK(\Omega)$ component of $\omega_t$ through the following theorem:\vspace{.25cm}\\
\textbf{Theorem 4.} If $\Omega_t$ is a disjoint union of solid tori then the orthogonal projection of the vorticity $\omega_t$ on $HK(\Omega_t)$ has constant fluxes at each time $t$ that is
\begin{equation}
\Phi_{it}=\Phi_{i0}=\int_{\Sigma_{0t}}\omega_{0}\cdot dS_0\,\,\,\,\,\,\,\,\,\,(i=1,\ldots,n).
\end{equation}\vspace{.25cm}\\
\emph{Proof:} Since $\omega_t$ belongs to the image of the curl operator, Eq.~(56) shows that it splits as
\begin{equation}\nonumber
\omega_t=\omega_t^{HK}+\omega_t^{FK}+\omega_t^{CG},
\end{equation}
where upper indices $HK$,$FK$,$CG$ are understood through Eq.~(56). On the other hand $\omega_t$ is tangent to the boundary and $\omega_t^{FK}$ and $\omega_t^{HK}$ are also tangent to the boundary and so must be $\omega_t^{CG}$, but due to the uniqueness of solutions of the Laplace equation with the Neumann boundary condition, $\omega_t^{CG}=0$. In addition $\omega_t^{FK}$ has zero flux on all interior surfaces such as $\Sigma_{it}$'s. Thus the flux $\Phi_{it}$ of $\omega_t^{HK}$ is equal to the flux of $\omega_t$:
\begin{equation}
\Phi_{it}=\int_{\Sigma_{it}}\omega_{it}\cdot dS_t.
\end{equation}
The last integral is a constant of the motion by a direct application of Eq.~(23) because each surface $\Sigma_{it}$ lies entirely inside the fluid. An alternative proof for these flux conservations is provided by the Kelvin circulation theorem as seen below. This integral according to the Stokes' theorem converts to the line integral on $\partial\Sigma_{it}$ all frozen into the fluid motion and thus:
\begin{equation}\nonumber
\frac{d}{dt}\int_{\Sigma_{it}}\omega_{it}\cdot dS_t=\frac{d}{dt}\oint_{\partial\Sigma_{it}}u_t\cdot dl_t,
\end{equation}
and since $dl_t$ is advected by the fluid and so the above derivative equals
\begin{equation}\nonumber
\oint_{\partial\Sigma_{it}}\frac{Du_t}{Dt}\cdot dl_t+\oint_{\partial\Sigma_{it}}u_t\cdot \frac{D}{Dt}dl_t.
\end{equation}
From Eqs.~(22) and (43) one can easily see that the two above integral reduce to the integrals of gradient fields along the loop $\partial\Sigma_{it}$ and therefore vanish and this completes the proof. Thus we obtained
\begin{equation}
\omega_t^{HK}=\sum_{i=1}^{n}\Phi_{i0}f_{it},
\end{equation}
$\blacksquare$\vspace{2.5mm}\\
The proof for the conservation of the flux $\Phi_{it}$ in Eq.~(69) implies a similar proof for this fact that for each $i$, $f_{it}/\rho_{t}$ is the pushforward of $f_{i0}/\rho_{0}$ because the flux of $f_{i0}$ on $\Sigma_{i0}$ is equal to the flux of $f_{it}$ on $\Sigma_{it}$ and $\Sigma_{it}$ is advected by the fluid motion. Thus we find
\begin{equation}\nonumber\
f_{it}=\rho_{t}(h_{t})_{*}(\frac{f_{i0}}{\rho_{0}}).
\end{equation}

It is seen in Eq.~(54) that the three subspaces of the righthand side are indeed gradient spaces and we can denote all of them with the general name $G(\Omega)$ subspace and simplify this equation to
\begin{equation}\nonumber
\Gamma(\Omega)= FK(\Omega)\oplus HK(\Omega)\oplus G(\Omega),
\end{equation}
and so all curl-free vector fields according to Eq.~(55) splits into $HK(\Omega)\oplus G(\Omega)$. Now use this in Eq.~(49) in which $u_t-BS(\omega_t)$ is obviously curl-free and thus we can write
\begin{equation}
u_t-BS(\omega_t)=\Upsilon_t +\nabla_{y_{t}}\varphi_t,
\end{equation}
where $\Upsilon_t\in HK(\Omega_t)$ and $\nabla_{y_{t}}\varphi_t\in G(\Omega_t)$ and thus Eq.~(49) by the use of (70) yields
\begin{equation}
\Delta \mathcal{H}(u_t)=\int_{\Omega_t}\Upsilon_t\cdot\omega_{t} d^{3}y_{t}=<\Upsilon_t\mid \omega_t^{HK}>=\sum_{i=1}^{n}\Phi_{i0}\kappa_{it}=\text{constant},
\end{equation}
where $ \kappa_{it}$'s are the circulations of $\Upsilon_{t}$ that is
\begin{equation}
\Upsilon_{t}=\sum_{i=1}^{n}\kappa_{it}l_{it}.
\end{equation}
The harmonic knot field $\Upsilon_t$ therefore has zero curl and divergence with respect to $y_t$ and tangent to $\partial\Omega_t$ at each time $t$:
\begin{equation}
\Upsilon_t\cdot ds_{t}=0.
\end{equation}
By the hypothesis $\Upsilon_t$ can not be a gradient field and of course when $\Omega_t$ is simply connected $HK(\Omega_t)=0$ and so $\Upsilon_t$ is identically zero. However, in general Eq.~(72) implies that at each time $t$, $\Upsilon_{t}$ lies in an $n-1$ dimensional hyperplane and specially when $ \Delta \mathcal{H}=0 $, $\Upsilon_t$ belongs to an  $n-1$ dimensional subspace of $HK(\Omega_t)$. To summarize, we conclude the following theorem.\vspace{2.5mm}\\
\textbf{Theorem 5.} If the domain $\Omega_0$ is simply connected so will be $\Omega_t$ and the Biot-Savart helicity equals the physical helicity and when these two helicities coincide, $\Omega_{t}$ is not necessarily simply connected but $\Upsilon_{t}$ belongs to a subspace of $HK(\Omega_t)$ with codimension one. In general $\Upsilon_{t}$ always lies in a geometrical hyperplane with codimension one.\vspace{2.5mm}\\
The above theorem clearly shows an algebraic structure for $\Delta \mathcal{H}$ and restricts $\Upsilon_{t}$ by one degree.

An alternative value for $\Delta \mathcal{H}$ is obtained using the vector identity written just before Eq.~(35) for the pair $\Upsilon_{t}$ and $u_{t}$, in Eq.~(72) and if we notice that $\Upsilon_{t}$ is curl-free after converting the volume integral into the surface integral we find
\begin{equation}\nonumber
\Delta \mathcal{H}(u_t)=\int_{\partial\Omega_t}(\Upsilon_t\times u_{t})\cdot ds_{t}.
\end{equation}
Substitute $\Upsilon_{t}$ from (71) into the above equation to obtain
\begin{equation}\nonumber
\Delta \mathcal{H}(u_t)=\int_{\partial\Omega_t}(u_{t}\times BS(\omega_{t}))\cdot ds_{t}+\int_{\partial\Omega_t}(u_{t}\times\nabla_{y_{t}}\varphi_{t})\cdot ds_{t}.
\end{equation}
Convert the second integral into the volume integral and then use again the vector identity written just before Eq.~(35) for the pair $u_{t}$ and $\nabla_{y_{t}}\varphi_{t}$. Then the second integral becomes
\begin{equation}\nonumber
\int_{\Omega_t}(\omega_{t}\cdot\nabla_{y_{t}}\varphi_{t}) d^{3}y_{t}=\int_{\partial\Omega_t}\varphi_{t}\omega_{t}\cdot ds_{t}=0.
\end{equation}
Hence we can calculate $\Delta \mathcal{H}$ in terms of $u_{t}$ and write
\begin{equation}
\Delta \mathcal{H}(u_t)=\int_{\partial\Omega_t}[u_{t}\times BS(\nabla_{y_{t}}\times u_{t})]\cdot ds_{t}.
\end{equation}

Now assume $\Omega_t$ is not simply connected and consider a closed curve $C_{it}$ defined like that defined before Eq.~(59), so each $C_{it}$ is generally a knot following its corresponding handlebody of the relevant solid torus. Therefore any Seifert surface $D_{it}$ whose boundary is $C_{it}$ can not be contained in $\Omega_t$ and from Eqs.~(59) and (71) it follows
\begin{equation}
\oint_{C}u_t\cdot dl_{t}=\oint_{C}BS(\omega_t)\cdot dl_t+\oint_{C}\Upsilon_t\cdot dl_t.
\end{equation}
As mentioned before the vector field $BS(\omega_t)$ always extends to $\mathbb{R}^3$ and if $u_t$ is assumed to extend too, so will be $\Upsilon_t$. The extended $\Upsilon_t$ is not curl-free outside  $\Omega_t$ because otherwise it is a curl-free vector field on $\mathbb{R}^3$ and so is a gradient field. Application of Stokes' theorem in the above equation leads to
\begin{equation}
\oint_{C_{it}}u_t\cdot dl_{t}=\int_{D_{it}\cap\Omega_t}\omega_t\cdot dS+\int_{D_{it}\cap\Omega_t}\nabla_{y_{t}}\times\Upsilon_t\cdot dS,
\end{equation}
which clearly demonstrates the effect of external flows on $\Omega_t$.

The harmonic knot field $\Upsilon_t$ is not necessarily frozen into the fluid motion but rather it remains always tangent to $\partial\Omega_{t}$ (Eq.~(74)) and without any divergence and curl. The conservation of $\Delta \mathcal{H}(u_t)$  according to Eq.~(72) and using $dm=\rho_t d^{3}y_{t}$ gives
\begin{equation}\nonumber
\int_{\Omega_t}\frac{\omega_t}{\rho_{t}}\cdot\Upsilon_{t}\rho_td^{3}y_{t}=\int_{\Omega_0}\frac{\omega_0}{\rho_{0}}\cdot\Upsilon_{0}\rho_0d^{3}x,
\end{equation}
in which the lower index zero denotes the initial value (at $t=0$). Assume $\Lambda_{ij}^{t}=\partial y_{ti}/\partial x_{j}$ is the Jacobi matrix whose determinant was already denoted by $J_{t}$. With this notation Eq.~(46) becomes $\frac{\omega_t}{\rho_t}=\Lambda^{t}\frac{\omega_0}{\rho_0}$ which its substitution into the above equation after simplification yields a new form for this conservation as follows:
\begin{equation}
\langle\omega_0\mid\tilde{\Lambda}^{t}\Upsilon_{t}-\Upsilon_{0}\rangle=0,
\end{equation}
where $\tilde{\Lambda}^{t}$ is the transpose of $\Lambda^{t}$.

%%%%%%%%%%%%%%%%%%%%%%%%%%%%%%%%%%%%%%%%%%%%%%%%%%%%%%%%%%%%%%%%%%%%%%%%%%%%%%%%%%%%%%%%%%%%
\section{Helicity in magnetohydrodynamics}

As cited in the Introduction the concept of helicity discovered in ideal vortex dynamics as well as ideal magnetohydrodynamics (MHD) both of which were found almost at the same time. Ideal MHD studies the dynamics of electrically conducting magnetized fluids including systems such as ideal plasmas, liquid metals, and electrolytes. Since an MHD system contains a fluid, the continuity equation (42) is one of its governing equations while another equation is obtained by the combination of Faraday's equation and the ideal conductivity condition as:
\begin{equation}
(\frac{\partial B_{t}}{\partial t})_{y_{t}}=\nabla_{y_t}\times(u_{t}\times B_{t}),
\end{equation}
where $B_t$ is the magnetic field in the compact closed domain $\Omega_t\subset\mathbb{R}^{3}$ with smooth boundary being advected by the fluid motion exactly similar to what considered before and the magnetic field is also tangent to the boundary of the domain. The above equation combined with the continuity equation yields the frozenness of magnetic field quite similar to Eqs.~(45) and (46) with $\omega_t$ replaced by $B_t$. The last governing equation is the equation of motion which contains the physical dynamics of the fluid:
\begin{equation}
\frac{D u_t}{D t}=-\nabla_{y_{t}} w_t+\frac{1}{c}j_t\times B_t,
\end{equation}
where $w_t$ is the thermodynamical enthalpy per unit mass, $j_t$ is the electrical current density vector and $c$ is the speed of light in vacuum and the fluid is considered to be compressible and isentropic similar to previous sections.

The above MHD equations admit the conservation of three different helicities. The first helicity is the Biot-Savart helicity which is a direct consequence of the magnetic frozenness similar to what obtained for the vorticity (Theorem 1):
\begin{equation}
\mathcal{H}_{BS}(B_t)=\int_{\Omega_t}B_t\cdot BS(B_t) d^3y_{t}=\text{constant}.
\end{equation}
The second helicity $\mathcal{H}_{M}$ is ``magnetic potential helicity'' related to the magnetic vector potential. To derive this helicity first we justify the existence of a magnetic vector potential $A_t$ such that $B_t=\nabla_{y_t}\times A_t$. The magnetic field $B_t$ here is the curl of $BS(B_t)$ and hence it belongs to $image(curl)$ which demonstrates the existence of the magnetic potential. Alternatively we may use the following argument: $B_t$ is divergence-free and thus belongs to $ker(div)=FK\oplus HK\oplus CG\oplus HG$ according to the Hodge theorem. The gradient part of this kernel consisting of $CG\oplus HG$ vanishes since $B_t$ is tangent to the boundary and so $B_t$ splits as
\begin{equation}
B_t=B^{FK}_t+B^{HK}_t,
\end{equation}
and immediately because of Eq.~(56) the magnetic field lies in $image(curl)$ and this justifies the existence of the vector potential $A_t$. Moreover, since physically magnetic field appears either due to electrical currents (through the Biot-Savart operator) or the Faraday induction law, the vector potential always exists. Substituting $B_t=\nabla_{y_t}\times A_t$ into Eq.~(79) and removing the curl operator because of Eq.~(55) we find
\begin{equation}
\frac{\partial A_t}{\partial t}=u_t\times \nabla_{y_t}\times A_t+\nabla_{y_t}\phi_t+\Pi_t,
\end{equation}
where $\nabla_{y_t}\phi_t\in G(\Omega_t)$ and $\Pi_t\in HK(\Omega_t)$ which yet can be arbitrarily selected. The $i$th component of the above equation can be written as
\begin{equation}
\frac{D A_{ti}}{Dt}=-A_{tj} \frac{\partial u_{tj}}{\partial y_{ti}}+\frac{\partial \tilde{\phi}_t}{\partial y_{ti}}+\Pi_{ti},
\end{equation}
where $\tilde{\phi}_t=\phi_t+u_t\cdot A_{t}$. Now define $\mathcal{H}_{M}$ as
\begin{equation}
\mathcal{H}_{M}(A_t)=\int_{\Omega_t}A_t\cdot B_t dy_{t}^3.
\end{equation}
The time derivative of this helicity is
\begin{equation}\nonumber
\frac{d}{dt}\mathcal{H}_{M}(A_t)=\int_{\Omega_t}\left(\frac{DA_t}{Dt}\cdot \frac{B_t}{\rho_t}+A_t\cdot\frac{D}{Dt}(\frac{B_t}{\rho_t})\right)\rho_td^3y_{t}.
\end{equation}
Using Eq.~(45) with $\omega_t$ replaced by $B_t$ and Eq.~(84) in the above equation yields
\begin{equation}
\frac{d}{dt}\mathcal{H}_{M}(A_t)=<\Pi_t\mid B_t>=<\Pi_t\mid B^{HK}_t>.
\end{equation}
Exactly the same argument which led to Theorem 4 and Eq.~(70) is valid for the magnetic field so we have
\begin{equation}
B_t^{HK}=\sum_{i=1}^{n}\Phi^{M}_{i0}f_{it},
\end{equation}
where $\Phi^{M}_{i0}=\int_{\Sigma_{it}}B_t\cdot ds_t$ are constant magnetic fluxes. The harmonic knot field $\Pi_t$ is expanded in terms of base vectors $\{l_{1t},\ldots l_{nt}\}$ as
\begin{equation}
\Pi_{t}=\sum_{i=1}^{n}\kappa^{M}_{it}l_{it},
\end{equation}
by which Eq.~(86) becomes
\begin{equation}
\frac{d}{dt}\mathcal{H}_{M}(A_t)=\sum_{i=1}^{n}\Phi^{M}_{i0}\kappa^{M}_{it}.
\end{equation}
The conservation of magnetic potential helicity directly comes from Eq.~(86) if
\begin{equation}
\langle\Pi_t\mid B_t\rangle=\langle\Pi_t\mid B^{HK}_t\rangle=0,
\end{equation}
for any compact smooth-boundary domain $\Omega_t$ and in particular if $\Omega_t$ is a disjoint union of solid tori Eq.~(90) according to (89) becomes
\begin{equation}
\sum_{i=1}^{n}\Phi^{M}_{i0}\kappa^{M}_{it}=0,
\end{equation}
which implies that $\Pi_t$ must belong to an $n-1$ dimensional subspace of $HK(\Omega_t)$. Hence, we have obtained:\vspace{2.5mm}\\
\textbf{Theorem 6.} For an arbitrary compact smooth-boundary domain $\Omega_t\subset \mathbb{R}^{3}$ occupied by an ideal conducting magnetized fluid, the magnetic vector potential $A_t$ satisfies Eq.~(83) and if Eq.~(90) holds, then the magnetic potential helicity $\mathcal{H}_{M}$ is constant and if $\Omega_t$ consists of disjoint solid tori $\Pi_t$ lies in a subspace with codimension 1. \vspace{2.5mm}\\
The conservation of this magnetic potential helicity of course is not a new result and as mentioned in the Introduction, it has been known for a long time. However this helicity was defined with $\Pi_t\equiv 0$ which is only a special case of Theorem 6 or it happens when $\Omega_t$  is simply connected. The difference between the magnetic potential and Biot-Savart helicities is again a constant which can be treated exactly like Theorem 5 and related results and so we ignore them.

The third conserved helicity is the well known ``cross helicity" $\mathcal{H}_{C}$ defined as
\begin{equation}
\mathcal{H}_{C}=\langle u_t\mid B_t\rangle=\text{constant},
\end{equation}
whose conservation can be easily seen when we write
\begin{equation}\nonumber
\frac{d}{dt}\mathcal{H}_{C}=\int_{\Omega_t}\left(\frac{Du_t}{Dt}\cdot \frac{B_t}{\rho_t}+u_t\cdot\frac{D}{Dt}(\frac{B_t}{\rho_t})\right)\rho_td^3y_{t}.
\end{equation}
Again using Eq.~(45) with $\omega_t$ replaced by $B_t$ into the second integral above, one finds that this integral is zero and the substitution of Eq.~(80) into the first integral, it is found to vanish too. It should be noted that according to Eq.~(80) the vorticity $\omega_t$ is no longer frozen into the fluid unlike to the hydrodynamical equation (43). However for a better understanding of this helicity suppose $\Omega_t$ is simply connected while $B_t$ and $\omega_t$ at each time $t$ are zero except on two narrow closed linked tubes $C_{B_{t}}$ and $C_{\omega_{t}}$ respectively. In this case it is not difficult to observe that
\begin{equation}
\mathcal{H}_{C}=Link(C_{B_{t}},C_{\omega_{t}})\Phi_{B_t}\Phi_{\omega_t}=\text{constant},
\end{equation}
where $\Phi_{B_t}$ and $\Phi_{\omega_t}$ are the fluxes of $B_t$ and $\omega_t$ respectively on their own tubes. The frozenness of $B_t$ guaranties the conservation of $\Phi_{B_t}$ and thus we find
\begin{equation}
Link(C_{B_{t}},C_{\omega_{t}})\Phi_{\omega_t}=\text{constant}.
\end{equation}
The magnetic tube will not be destroyed because of the frozenness while the vortex tube may diverge but until it keeps to be a narrow tube with a preserved linking number with the magnetic tube, Eq.~(94) gives the conservation of $\Phi_{\omega_t}$.

The first (Biot-Savart) helicity as we saw was conserved only due to the frozenness of $B_t$ which is not directly related to the equation of motion (80). The second (magnetic potential) helicity also was conserved due to the frozenness of the magnetic field and so these two helicities are of the same type and it has sense to compare them and discuss about their difference although it is not so interesting and probably has no new information. On the other hand the constancy of the third (cross) helicity is related to the equation of motion as well as the frozenness of $B_t$ and so this helicity is in different type with respect to the first and second helicities. Hence, the difference between the cross helicity and magnetic potential helicity is physically more interesting as it compares two constants steaming from two physical equations.
\begin{equation}
\mathcal{H}_{C}-\alpha \mathcal{H}_{M}(A_t)=\langle u_t-\alpha A_t\mid B_t\rangle=\text{constant},
\end{equation}
where $\alpha$ is a constant in order to equalize the physical units of the two helicities. From Eq.~(82) $B_t$ splits into two parts and so only these two parts of $u_t-\alpha A_t$ contribute to the above constant.

%%%%%%%%%%%%%%%%%%%%%%%%%%%%%%%%%%%%%%%%%%%%%%%%%%%%%%%%%%%%%%%%%%%%%%%%%%%%%%%%%%%%%%%%%%%%%%%%%%%%%%%%%%%%%%%%%%%%%%%%%%%%%%%%%%%%%%%%%%%%%%%%%%%
\section{Summary and remarks}

The present note was devoted to the old fundamental debate of topological invariants of an ideal flow during the motion. Such flows leave a number of topological properties invariant that certainly relate to some constants of motion among which helicity may be the simplest and still the richest conserved quantity containing valuable information. The well known model of a family of smooth one parameter orientation-preserving diffeomorphisms on a compact smooth boundary domain in the 3-space was used to describe an isentropic fluid motion. However unlike to the previous investigations, here the diffeomorphisms were not restricted to be volume-preserving (corresponding to an incompressible flow) and instead, they were assumed to be smooth and nonsingular and consequently preserve a finite measure (modeling the fluid mass) to display the compressibility of the flow. It was seen that the frozenness of the vortex lines into the fluid motion means that the vorticity filed divided by the fluid density at any time is the pushforward of its initial field and moreover the properties of being divergence-free and tangent to the boundary was established to be preserved at all times. Such a situation allowed us to define a Biot-Savart helicity whose conservation was directly demonstrated while the energy of vortex field (enstrophy) is no longer a constant of motion and so its time variation was calculated to show that it has the same form for both compressible and incompressible flows. In two special cases the Biot-Savart helicity is significantly simplified and is closed to the energy: when vorticity is an eigenvector of the curl operator and when it is an eigenvector of the Biot-Savart operator which the latter case has been widely studied in the literature to find upper bounds for the helicity. It was shortly mentioned here the situation were these two cases give the same helicity and energy although a detailed precise analysis looks to have sense.

Next it was noticed to the physical equations of an ideal isentropic fluid namely, the continuity equation and the curl of the equation of motion which led to the frozenness of the vortex field or equivalently the pushforwardness of the vorticity divided by the fluid density. This established the conservation of the Biot-Savart helicity as a consequence of the \emph{curl} of the equation of motion and the equation of motion itself is not necessary for this helicity. On the other hand, the physical helicity conservation was reviewed as a result of the equation of motion (without taking the curl) with the emphasis on the fact that the values of the Biot-Savart and physical helicities are generally different which coincide when the fluid domain is simply connected. Especially when the fluid domain is not simply connected and the fluid velocity can be extended to the whole of the 3-space, the circulation of the fluid velocity on a non trivial loop is generally different from the circulation of the Biot-Savart of the vorticity. To understand the difference between these two helicities the Hodge decomposition theorem for domains in 3-space was used and it was seen that the harmonic part of the vorticity field at any time plays the main role in the value of the helicity difference. The harmonic knot subspace of the fluid domain at each time is known to be isomorphic to an Euclidian space with dimension equal to the sum of the genuses of all components of the boundary of the fluid domain. However it was found here that if the fluid domain consists of finite number of disjoint solid tori, two sets of base functions for the harmonic subspace are available: The set of functions each having unit flux on a definite cross section cutting a special handlebody, and alternatively the set of functions each having unit circulation on the loop following a special handlebody. Moreover it was shown that each function in the first basis is orthogonal to its corresponding function in the other basis and a harmonic knot field is uniquely determined by its fluxes or its circulations. The fluid velocity minus the Biot-Savart of the fluid vorticity defines a curl-free vector field whose harmonic part was also important in constructing the helicity difference which was shown to stay in a hyperplane with codimension 1 when it is expanded in terms of the second base functions above and this hyperplane reduces to a supbspace with codimension 1 if the two helicities coincide. This property came from the obtained result that the expansion of the harmonic part of the vorticity in terms of the first base functions has constant fluxes equal to the fluxes of the vorticity. It was also mentioned briefly that the Kelvin circulation theorem is in general a consequence of the equation of motion and is also derived from the vortex frozenness when the fluid domain is simply connected.

In addition to vortex dynamics, magnetohydrodynamics was also observed to manifest the classical helicity conservation in nature although later on various appearances of the subject happened in relativistic vortex dynamics and MHD \cite{eshraghi} as well as in modern theoretical and applied physics \cite{mahajan1}-\cite{keida}. Here it was reviewed that the continuity equation and the Faraday law combined with the infinite electric conductivity produces the frozenness of the magnetic field into the fluid motion directly leading to the magnetic Biot-Savart helicity conservation in which the equation of motion was not used. Therefore when the equation of motion was taken into account, two independent helicity conservations arose, the magnetic potential and the cross helicities. The traditional version of the magnetic potential helicity had been discovered mainly for a simply connected fluid domain and here it was generalized to the case where a harmonic knot field naturally appears in the dynamics of the magnetic potential. It was shown that if this harmonic knot is orthogonal to the harmonic part of the magnetic field, this generalized magnetic potential helicity will be constant. In the special case when the fluid domain is made of disjoint solid tori, the harmonic part of the magnetic field is represented by a set of constant fluxes across handlebodies and thus the magnetic potential helicity conservation will be held if the harmonic knot field appeared in the magnetic potential dynamics lies in a subspace of codimension 1. The cross helicity conservation (whose proof was reviewed here for the completeness) is perhaps the most natural invariant since it directly deals with pure physical observables namely, fluid velocity and magnetic filed. However it was seen that when each of vortex and magnetic fields restrict to narrow linked tubes, the cross helicity is proportional to their linking number and moreover the flux of the vortex tube is preserved up to its existence without intersecting the magnetic tube.

%%%%%%%%%%%%%%%%%%%%%%%%%%%%%%%%%%%%%%%%%%%%%%%%%%%%%%%%%%%%%%%%%%%%%%%%%%%%%%%%%%%%%%%%%%%%%%%%%%%%%%%%%%%%


\begin{thebibliography}{99}
\footnotesize

\bibitem{moffatt1}
H.~K.~Moffatt, Some remarks on topological fluid mechanics In, {\it An introduction to the geometry and topology of fluid flows}, R.~l.~ Ricca (editor), pp 3-10 (Kluwer Academic, 2001) .
\bibitem{helm}
L.~F.~Helmholtz, \"{U}ber integral der hydrodynamischen Gleichungen welche den Wirbelbewegungen entsprechen, {\it journal f\"{u}r reine und angewandte Mathematik.} {\bf 55}, pp 25-55 (1858).
\bibitem{kelvin}
W.~Thomson, On vortex motion, {\it Trans. R. Soc. Edinb.} {\bf 25}, pp 217-260 (1869).
\bibitem{poincare}
H.~Poincar\'{e}, {\it Sur la th\'{e}orie des tourbillion.} (Gauthier-Villars, 1893).
\bibitem{chandra}
S.~Chandrasekhar, J.~Woltjer, On force-free magnetic fields. {\it Proceedings of the National Acad. Sci.} {\bf 44}, pp 285-289 (1958).
\bibitem{woltjer1}
J.~Woltjer, A theorem on force-free magnetic fields, {\it Proceedings of the National Acad. Sci. USA} {\bf 44}, pp 489-491 (1958).
\bibitem{woltjer2}
J.~Woltjer, On Hydromagnetic equilibrium, {\it Proceedings of the National Acad. Sci. USA} {\bf 44}, pp 833-841 (1958).
\bibitem{moffat2}
H.~K.~Moffatt, The degree of knottedness of tangled vortex lines, {\it J. Fluid. Mech.} {\bf 36}, pp 117-129 (1969).
\bibitem{moffat3}
H.~K.~Moffatt, Fluid Mechanics, Topology, Cusp Singularity, and Related Matters Lecture delivered to the first ``Seminaire International de ÍInstitut de Mécanique de Grenoble", 19-21 May (1992).
\bibitem{gauss}
C.~F.~Gauss, Integral formula for linking number, in {\it Zur mathematischen Theorie de Electrodynamische Wirkungen}, Collected Works, 2nd ed {\bf 5}, pp 605 (Koniglichen Gesellschaft des Wissenschaften, Gottingen, 1833).
\bibitem{caluga1}
G.~C\u{a}lug\u{a}reanu, L'int\'{e}gral de Gauss er l'analyse des noeuds tridimensionnels, {\it Rev. Math. Pures Appl.} {\bf 4}, pp 5-20 (1959)
\bibitem{caluga2}
G.~C\u{a}lug\u{a}reanu, Sur les classes d'isotopie des noeuds tridimensionnels et leurs invariant, {\it Czechoslovak Math.} J.~T. {\bf 11}, pp 588-625 (1961).
\bibitem{caluga3}
G.~C\u{a}lug\u{a}reanu, Sur les enlacements tridimensionnels des courbes ferm\'{e}es, {\it Comm. Acad. R.P. Romine}, {\bf 11} pp 829-832 (1961).
\bibitem{arnold1}
V.I~Arnold, The asymptotic Hopf invariant and its applications, {\it Proceedings of Summer School in Differential Equations} at Dilizhan, 1973 (1974),
Erevan: Armenian Academy of Science (in Russian); English transl. {\it Sel. Math. Sov.} {\bf 5}, pp 327-345 (1986).
\bibitem{arnold2}
V.I~Arnold, B.A~Khesin, {\it Topological methods in hydrodynamics} (Springer-verlag, New York, 1998).
\bibitem{khesin}
B.~A.~Khesin, Topological Fluid Dynamics, {\it Notices of American Math. Soc.} {\bf 52}, pp 9-19 (2005).
\bibitem{gambaudo1}
J.-M~Gambaudo, Knots, flows and fluids, {\it Dynamique des difféomorphismes conservatifs des surfaces: un point de vue topologique} No. 21, pp 53–103 (Panoramas Et Syntheses, 2006).
\bibitem{cantarella1}
J.~Cantarella, D.~DeTurck, H.~Gluck, and M.~Teytel, Isoperimetric problems for the helicity of vector fields and the Biot-Savart and curl operators, {\it J. Math. Phys.} {\bf 41}, pp 5615-5641 (2000).
\bibitem{cantarella2}
J.~Cantarella, D.~DeTurck,and H.~Gluck, The Biot-Savart operator for application to knot theory, fluid dynamics, and plasma physics,
{\it J. Math. Phys.} {\bf 42}, pp 876-905 (2001).
\bibitem{cantarella3}
J.~Cantarella, D.~DeTurck,and H.~Gluck, Upper Bounds for the Writhing of Knots and the Helicity of Vector Fields, {\it Studies in Advanced Mathematics} {\bf 24}, pp 1-21 (2001).
\bibitem{contreras}
G.~Contreras, and R.~Iturriaga, Average linking number, {\it Ergodic Theory Dynam. systems} {\bf 19}, pp 1425-1435 (1999).
\bibitem{gambaudo2}
J.-M.~Gambaudo, and É.~Ghys, Enlacements asymptotiques, {\it Topology} {\bf 36}, pp 1355-1379 (1997).
\bibitem{cantarella4}
J.~Cantarella, D.~DeTurck,and H.~Gluck, Vector calculus and the topology of domains in 3-space, {\it American Mathematical Monthly} {\bf 109}, pp 409-442 (2002).
\bibitem{Chorin}
A.~J.~Chorin and J.E.~Marsden, {\it A Mathematical Introduction to Fluid Mechanics} (Springer-Verlag, New York, 1979).
\bibitem{warner}
F.~W.~Warner, {\it Foundations of Differentiable Manifolds and Lie Groups} (Springer-Verlag, New York, 1983).
\bibitem{eshraghi}
H.~Eshraghi, On the vortex dynamics in fully relativistic plasmas, {\it Phys. Plasmas} {\bf10}, pp 3577-3583 (2003).
\bibitem{mahajan1}
B.~A.~Bambah, S.~M.~Mahajan, C.~Mukku, Yang-Mills Magnetofluid Unification, {\it Phys. Rev. Lett.} {\bf97}, p 072301 (2006).
\bibitem{mahajan2}
S.~M.~Mahajan, F.~A.~Asenjo, Vortical dynamics of spinning quantum plasmas: Helicity conservation, {\it Phys. Rev. Lett.} {\bf107}, p 195003 (2011).
\bibitem{keida}
H.~Kedia, I.~Bialynicki-Birula, D.~Peralta-Salas, W.~T.M.~Irvine, Tying knots in light fields, {\it Phys. Rev. Lett.} {\bf111}, p 150404 (2013).
\bibitem{cantarella5}
J.~Cantarella, J.~Parsley, A new cohomological formula for helicity in $\mathbb{R}^{2k+1}$ reveals the effect of a diffeomorphism on helicity, {\it J. Geom. Phys.} {\bf60:9}, p 1127-1155 (2010).
\bibitem{Cantarella6}
J.~Cantarella, A General Cross-Helicity Formula, {\it Proceedings of the Royal Society, Series A} {\bf456}, p 2771-2779 (2000).
%%%%%%%%%%%%%%%%%%%%%%%%%%%%%%%%%%%%%%%%%%%%%%%%%%%%%%%%%%%%%%%%%%%%%%%%%%%%%%%%%%%%%%%%%%%%%%%%%%%%%%%%%%%%%%%%%%%%%%%%%%%%%%%%%%%%%%%%%%%%%%%%%%%

\end{thebibliography}
\end{document}